\documentclass[twocolumn,floatfix,prl,showpacs]{revtex4}
\usepackage[dvips]{graphicx}
\usepackage{amsmath}
\usepackage{bm}
\usepackage{hyperref}

\usepackage[dvips]{color}

\begin{document}
    \title{Model of an exotic chiral superconducting phase in a graphene bilayer}

\author{Mir Vahid Hosseini and Malek Zareyan}

\affiliation{Department of Physics, Institute for Advanced Studies
in Basic Sciences (IASBS), Zanjan 45137-66731, Iran}

\begin{abstract}
We theoretically demonstrate the formation of a new type of
unconventional superconductivity in graphene materials, which
exhibits gapless property. The studied superconductivity is based
on an interlayer pairing of \textit{chiral} electrons in bilayer
graphene, which results in an exotic s-wave spin-triplet
condensate order with anomalous thermodynamic properties. These
include the possibility of a \textit{temperature-induced
condensation} causing an increase of the pairing gap with
increasing temperature, and an entropy of the stable
superconducting state which can be \textit{higher} than its
value in the normal state. Our study reveals the analogy of the
interlayer superconductivity in graphene materials to the color
superconductivity in dense quark matter and the gapless pairing
states in nuclear matter and ultra-cold atomic gases.
\end{abstract}

\maketitle

The microscopic Bardeen-Cooper-Schrieffer (BCS) theory of
superconductivity \cite{BCS}, is based on the formation of the so
called Cooper pairs between electrons with opposite spin and equal
but opposite momentum, when a weak attractive interaction exists
between the electrons. Despite first being applied specifically to
explain metallic superconductivity, the BCS pairing was discovered
later on to be a general effect which could appear in other areas
of physics (see Ref. \cite{BCSWil} for a recent overview). While
in conventional metallic superconductors an energy gap is opened
in the electronic excitation spectrum on the Fermi
surface\cite{Super1}, unconventional pairing states can be
realized in a wide range of fermionic systems in which
superconductivity coexists with the gapless spectrum of the normal
(N) phase. A great interest has been devoted to the BCS pairing in
composite systems of two (or more) types of fermions with
different Fermi surfaces, in which despite a nonvanishing pairing
gap (PG) or order parameter, there appears no energy gap on the Fermi
surface due to a separation of N and superconducting (S) phases in the momentum
space\cite{WilczekBP}. Gapless color superconductivity in
dense quark matter at low temperatures \cite{color1} and its analogous
gapless pairing states in nuclear matter \cite{NuclMatt} and
ultra-cold atomic gases with two types of atoms or identical atoms with different
spin states in an external magnetic (see \textit{e. g.} \cite{ColdAtoms}),
are famous examples of this kind.
\begin{figure}[t]
\begin{centering}\includegraphics[width=7.cm]{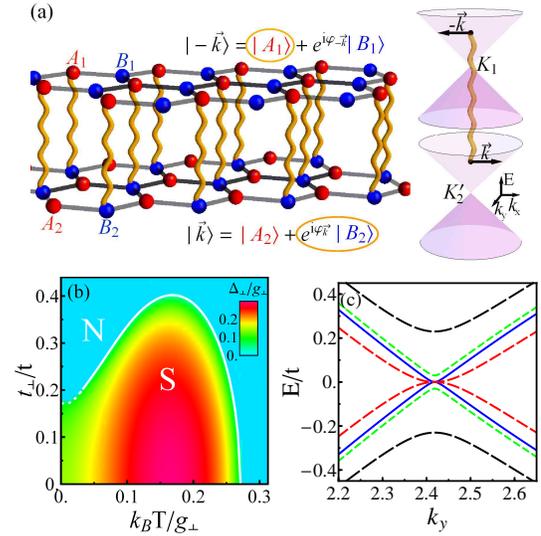}
\par\end{centering}
\caption{\label{hzfig1}{\small (a) Sketch of the bilayer graphene lattice
structure with the interlayer chiral superconducting correlations
(illustrated by wavy lines) between electrons of different types
of sublattices. Dirac cone bands of two monolayers
of graphene on the valleys $K_{1}$ and $K^{\prime}_{2}$ (right panel),
with two states of wave vector $|k\rangle$ and $|-k\rangle$ which are coherent
superpositions of the sublattice pseudospin states $|A_{1,2}\rangle$ and
$|B_{1,2}\rangle$. The interlayer pairing between
$|\pm k\rangle$ is asymmetrically partial, since only $|A_{1}\rangle$ and
$|B_{2}\rangle$ components are paired. (b) The mean field
phase diagram of the chiral superconductivity showing dependence
on the interlayer hopping energy $t_{\perp}$ and the temperature
$T$ for a pure bilayer graphene ($\mu=0$). The line of transition
from N to S phases is shown in which dashed and solid parts
indicate the first and the second order transitions, respectively.
Color plot of the pairing gap (PG) is also presented. (c) The resulted gapless
S band structure along $k_y$ for undoped bilayer graphene and for
$t_{\perp}=0.13t$.}}
\end{figure}
\par
The color superconductivity is the state of matter at highest densities
and low temperatures which can be reached at the interior
of dens neutron stars\cite{color1}. In this Letter we report on the
theoretical study of interlayer electronic pairing in pure bilayer
graphene, which explores the potential of this carbon-based
material to lend itself to an experimentally feasible study of the
relativistic gapless superconductivity. This may also simulate the
regime of color superconductivity. Graphene\cite{graphene1}, the
two dimensional solid of carbon atoms with honeycomb lattice structure,
has already been proven to exhibit a variety of compelling pseudo-relativistic and
quantum electrodynamics phenomena such as Klein tunneling
\cite{Klein}, Zitterbewugung
\cite{zitterbewegung} and universal light
absorption\cite{uLight}, to name a few. This is due to its unique
zero-gap electronic band structure with conically shaped
conduction and valence bands touching each other at the corners of
hexagonal first Brillouin zone, known as Dirac points (see
Fig.~\ref{hzfig1}a). An important aspect in graphene
is the connection between this specific band structure and the
sublattice pseudospin degree of freedom, which makes electrons
behave like two dimensional massless Dirac fermions with a
pseudo-relativistic chiral property\cite{graphene2}.
Here we find that in bilayer graphene, the interplay between the
interlayer pairing of electrons with the same sublattice chirality
and the asymmetric arrangement of the sublattices of the two
layers results in a gapless superconducting which posses an exotic
s-wave spin triplet symmetry of the order parameter and an anomalous
N-S phase diagram, shown in Fig.~\ref{hzfig1}b. We further find that the
entropy in the chiral S state can be higher than the entropy in N state. This
finding is in striking contrast to the common behavior of the entropy upon
a phase transition.
\par
We start with a qualitative description of the interlayer pairing
in bilayer graphene, which reveals the key features of the chiral
superconductivity and its anomalous properties. As is shown in
Fig.~\ref{hzfig1}a, in bilayer graphene two monolayers with $A_1$
and $B_1$ triangular sublattices in top layer and $A_2$ and $B_2$
triangular sublattices in bottom layer, are arranged according to
Bernal stacking in which every $A_1$ site of top layer lies
directly above a $B_2$ site of bottom layer. In the absence of
hopping between the layers($t_{\bot}=0$) the electronic states of
the layers are not mixed in N state. In the layer 1(2) the
electronic states lie in two Dirac-cone bands on the two
inequivalent $K_{1(2)}$ and $K^{\prime}_{1(2)}$ valleys (Dirac
points). In each layer, pseudospin chirality links the relative
amplitude of electron wave function (to be in the A or B
sublattices) to the direction of its momentum. Each momentum state in
Dirac cones of layers 1,2 $|k_{1,2}\rangle$, is a coherent superposition
of the two sublattice pseudospin states $|A_{1,2}\rangle$ and  $|B_{1,2}\rangle$
as $|k_{1,2}\rangle=|A_{1,2}\rangle+exp(i\varphi_{k_{1,2}})|B_{1,2}\rangle$, with the angles
$\varphi_k{_{1,2}}$ determining the directions of the wave vectors $k_{1,2}$.
The interlayer superconductivity will pair electrons in A1 and B2 sublattices,
but not those in A2 and B1 sublattices, which implies that the
pairing between the states $|k_{1}\rangle$ and $|k_{2}\rangle$ is asymmetrically partial
with respect to the pseudospin degree of freedom. Taking into account
the chirality, together with the requirement that the paired electrons have equal but opposite
momentum on the valleys $K_{1}$ and $K^{\prime}_{2}$, we obtain a
rough qualitative picture of N and S states configuration in the momentum space.
In each valley, only electrons of one sublattice becomes S and the other
sublattice remains in N state. This simultaneous existence of N and S
phases can give rise to the gapless property of the chiral
superconductivity for an undoped bilayer. Having a partial pairing at zero temperature $T=0$,
a thermal excitation can redistribute electrons in N component of a wave vector states
in new momentum states of coherent superposition of A and B states, where they have potential
for a partial pairing again. This corresponds to a temperature-induced condensation
process in which two primarily normal electrons are thermally excited to the coherent states,
where they can now be partially paired and absorbed into the chiral superconducting condensate.
By this process the pair amplitude can increase with increasing temperature.
\par
The above described qualitative picture can be supported by a mean
field calculation based on tight-binding theory of $\pi$ electrons
in graphene. In the absence of superconductivity, we consider the
following Hamiltonian \cite{BilayerTB} for bilayer graphene
\begin{eqnarray}
\label{HN} &H_0&=-\mu \sum_{\ell, \sigma ,i }n_{\ell ,i,\sigma}
-t\sum_{\ell,\sigma ,\langle i,j\rangle }(a_{\ell
,i,\sigma}^{+}b_{\ell ,j,\sigma}+h.c.)\\
&-&t_{\perp}\sum_{i}(a_{1,i,\sigma}^{+}b_{2,i,\sigma}+h.c.)\notag,
\end{eqnarray}
where $a_{\ell ,i,\sigma}$ ($b_{\ell ,i,\sigma}$) and
$a^{\dagger}_{\ell ,i,\sigma}$ ($b^{\dagger}_{\ell ,i,\sigma}$)
are the annihilation and creation operators of an electron in the
$i$th unit cell in sublattice A(B) and layer $\ell (=1,2)$;
$\sigma=\pm$ denotes the spin state of the electron; $n_{\ell
,i,\sigma}$ is the corresponding on-site particle density
operator. The intralayer nearest neighbor hopping energy $t\approx
3 eV$ determines the Fermi velocity in graphene as $v_F
\approx10^6 m/s$, and $t_\perp \approx0.4eV$ ($t_\perp/t \approx
0.13$) is the dominant interlayer hopping energy between the
nearest neighbors A1 and B2; the chemical potential $\mu$ can be
controlled by gate voltages. The interlayer superconductivity can
be introduced by adding the following potential to $H_0$
\begin{eqnarray}
V_{\perp}=-g_\perp \sum_{i}\sum_{\sigma,\sigma^{\prime}}a_{1, i,
\sigma}^{\dagger} a_{1, i, \sigma}b_{2, i,
\sigma^{\prime}}^{\dagger}b_{2, i, \sigma^{\prime}},
\label{Vp}
\end{eqnarray}
where $g_\perp$ is the S coupling energy \cite{BaskaranCC,grapheneSuperTheor}. Below, we will discuss the origin of this attractive potential. The interlayer Coulomb repulsion, competing with this attractive potential, is expected to be rather weak because of the interlayer spacing between the electrons. We note that the potential (\ref{Vp}) presents an on-site local interaction in the 2D
plane of the bilayer, which requires the pairing to have an
isotropic s-wave orbital symmetry. On the other hand the pairing
is anti-symmetric with respect to the pseudospin (i.e. sublattice)
degree of freedom. Pauli exclusion principle requires that the
total wave function, composed of the product of the orbital, spin and
pseudospin components, should be anti-symmetric under the
exchange of electrons. Therefore we conclude that the interaction
(\ref{Vp}) allows only an spin-triplet pairing. We find that the
interacting Hamiltonian $H=H_0+V_{\perp}$ can be decoupled by the
following spin-triplet order parameter
\begin{eqnarray}
\Delta_{i,\perp}=-g_{\perp}\langle a_{1, i, \downarrow}b_{2, i,
\uparrow}+a_{1, i, \uparrow}b_{2, i, \downarrow} \rangle,
\label{Order}
\end{eqnarray}%
which describes the interlayer pairing of electrons with the sublattices
$A_1$-$B_2$, the valleys $K_1$-$K_2^{\prime}$ (and $K_1^{\prime}$-$K_2$ ), and
the spins $\uparrow$-$\downarrow$ (and $\downarrow$-$\uparrow$).
It is worth to note that while the unconventional s-wave spin-triplet pairing is expected to exhibit an odd-frequency superconductivity\cite{Oddfrequency}, we have found that the order parameter (\ref{Order}) has an even parity in the frequency space. This is the result of an odd-pseudospin parity, as described above.
We obtain the excitation spectrum by replacing the
mean field order parameter (\ref{Order}) in $H=H_0+V_{\perp}$ and performing the
Fourier transformation of the Hamiltonian in the space of 2D wave
vector $\bf k$ of electrons. For zero doping $\mu=0$, we obtain a
simple expression for the spectrum $E_{\bf k}^{s l}$ which reads
\begin{eqnarray}
\alpha E_{\bf k}^{s l}=\frac{\alpha}{2} [\sqrt{(\Delta_\perp+s
t_\perp)^2+4\epsilon_{\bf k}^2}+l(\Delta_\perp+s t_\perp)],
\label{eigen}
\end{eqnarray}
where $l,s,\alpha=\pm$ indicate different branches of the
spectrum. Here $\epsilon_{\bf k}=\pm t|\gamma_{\bf k}|$ with
$\gamma^{\star}_{\bf k}=\sum_{\bf \delta}\exp{-i\bf k.
\bf\delta}$, with $\delta$ indicating the nearest neighbor
position vectors of graphene. The spectrum (\ref{eigen})
represents a gapless superconductivity as can be seen in
Fig.~\ref{hzfig1}c, in which different branches of $E_{\bf k}^{s
l}$ versus $k_y$ are plotted when $\mu=0$.
\begin{figure}
\begin{centering}\includegraphics[width=7.5cm]{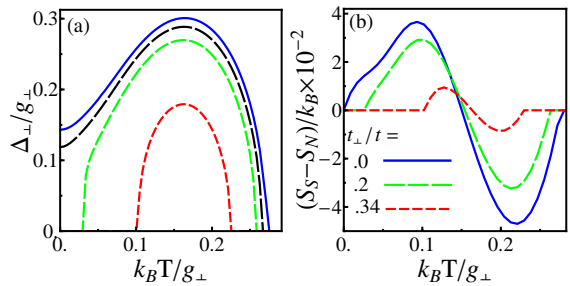}
\par\end{centering}
\caption{\label{hzfig2}{\small Exotic thermodynamic
properties of the chiral superconductivity. (a)  Pairing gap versus $T$ for
$t_{\perp}/t=0.0,0.13,0.2,0.34$ (from top to bottom, respectively).
(c) Difference in entropies of
N and S states versus $T$ for the same values of parameters as in (a).}}
\end{figure}
Now, having the excitation spectrum and using the standard many-body Green's
function method, we can obtain the self-consistent equation for
the pairing gap (PG) $\Delta_\perp$, as follows
\begin{eqnarray}\label{eq:final_result}
\Delta_{\perp}&=& \frac{g_\perp}{2}\sum_{\bf k, s, l}\frac{l
E_{\bf k}^{s l}}{\sqrt{(\Delta_\perp+st_\perp)^2+4 \epsilon_{\bf
k}^2}}\tanh(\frac{\beta E_{\bf k}^{s l}}{2}).
\end{eqnarray}
We can also obtain the thermodynamic potential $\Omega_S$ and the
entropy $S_S$ of S state from the following formulas, respectively,
\begin{eqnarray}\label{eq:Thermo}
\Omega_S=\frac{\Delta^2_\perp}{g_\perp}-\frac{1}{\beta}\sum_{\bf k
s l \alpha}\ln(1+\exp^{-\beta \alpha E_{\bf k}^{s l}}),
\end{eqnarray}
\begin{eqnarray}\label{eq:entropy}
S_S=-2k_B\sum_{\bf{k}, s, l}[f_{\bf k}^{s l}\ln f_{\bf k}^{s l}
+(1-f_{\bf k}^{s l})\ln(1-f_{\bf k}^{s l})],
\end{eqnarray}
where $f_{\bf k}^{s l}=[1+\exp(\beta E_{\bf k}^{s l})]^{-1}$ is
the Fermi-Dirac occupation factor with $\beta=1/k_BT$ ($k_B$ being
Boltzmann constant). The expressions for the thermodynamic
potential and entropy in N state follow from Eqs.
(\ref{eq:Thermo}) and (\ref{eq:entropy}) by taking
$\Delta_{\perp}=0$.
\begin{figure}
\begin{centering}\includegraphics[width=7.7cm]{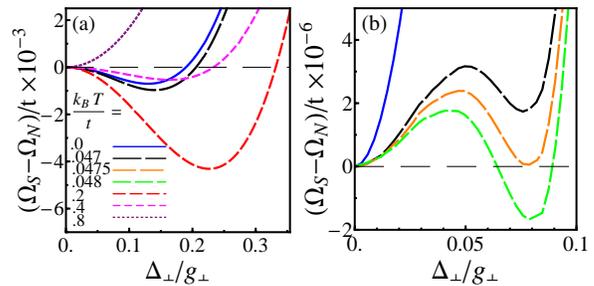}
\par\end{centering}
\caption{\label{hzfig3}{\small Stability of the exotic
chiral superconducting states. Difference in thermodynamic
potentials of N and S states versus the pairing gap calculated at
different temperatures for interlayer hopping energies $0.1t$ (a)
and $0.18t$ (b).}}
\end{figure}
\par
The temperature dependence of PG for different values of the
interlayer hopping energy can be obtained from a numerical
solution of Eq. (\ref{eq:final_result}). We have examined the
stability of the resulted S states by ensuring that any solution
of (\ref{eq:final_result}) presents indeed a global minimum of
$\Omega_S$ (see Fig.~\ref{hzfig3}). The resulting
phase diagram and $\Delta_\perp(T)$ is surprisingly anomalous,
as is presented in Fig.~\ref{hzfig1}b and Fig.~\ref{hzfig2}a, respectively,
for an undoped bilayer ($\mu=0$). For
$t_{\bot}=0$, the gap increases from its zero temperature value
with increasing temperature and reaches a maximum before showing a
BCS like decrease with $T$. As described above, this is the result
of thermal excitations of electrons from a N valence band into a
S conduction band, where they can form a condensate and contribute to the
PG. An interlayer hoping $t_{\perp}$ introduces a normal coupling
between the A1 and B2 sublattices and will weaken the
superconducting correlation and hence the interlayer PG. The
interplay between this effect and the temperature-induced increase
of the chiral condensation leads to the most amazing behavior of
PG. As is seen in Fig.~\ref{hzfig1}b and Fig.~\ref{hzfig2}a, it appears that for
$t_{\perp}$ exceeding some critical value, PG vanishes at a temperature range
starting from $T=0$, while it takes a sizable
value at higher temperatures.
\par
In the phase diagram of the chiral superconductivity in $t_{\bot}-T$
plane (Fig.~\ref{hzfig1}b) the color plot of PG is also shown. The
normal-superconducting (NS) transition is first
order at lower temperatures (the dashed part of a NS boundary),
and second order at higher temperatures (the solid part). This can
be seen from Fig.~\ref{hzfig3}, in which the potential difference
$\Omega_S-\Omega_N$ in N and S states is plotted versus
$\Delta_{\perp}$ for $\mu=0$ and two values of $t_{\bot}/t=0.1$
and $0.18$. While for $t_{\bot}/t=0.1$, the low
temperature NS transition is continuous, for $t_{\bot}/t=0.18$, it
happens discontinuously, as the stable solution $\Delta_{\perp}$
undergoes a jump from zero to a finite value at the transition.
\par
The anomalous behavior of $\Delta_{\perp}(T)$ is associated with
a very unusual behavior of the entropy, as is presented in
Fig.~\ref{hzfig2}b by plotting the entropy difference $S_S-S_N$ in
N and S states as a function of $T$ for different values of
$t_{\bot}$. The entropy difference undergoes a change of sign at
the temperature of maximum PG, which implies that the entropy of
the S state is higher than that of the N state, in the temperature
interval where PG increases with $T$. This unusual finding can
also be understood in terms of the above described
temperature-induced condensation process.
Similar to the above found exotic behavior of PG and the entropy
are also predicted to exists in the pairing state of asymmetric nuclear
matter \cite{NuclMatt}, and gapless color superconducting
phase of dens quark matter in which correlations between quarks
with three different colors and three different flavors results in
a variety of pairing patterns \cite{color1,color2}. A similar to
colored pairing of quarks can be simulated by the interlayer chiral
superconductivity in trilayer graphene with two different ABA and ABC
stackings \cite{TrilayerBS}, which provides pairing
between electrons whose states lie in three distinct Dirac cones
with opposite sublattice pseudospins.
\par
Finally, let us comment on the experimental realization of our predicted
exotic superconductivity. The possibility of an intrinsic
superconductivity, with plasmon or phonon mediated pairing interactions,
has been proposed in graphene coated with a metal \cite{grapheneSuperTheor}. For
a coated bilayer graphene the formation of the S state is expected to be closely similar to that
of the graphite intercalated with alkaline metals, for which a critical
temperature up to 11.5 K has been reported (see \textit{e. g.} \cite{graphiteSuper}). This can be
served as an estimation of the critical temperature (and essentially the coupling energy
$g_{\perp}$) for the interlayer coupling. Alternatively, the superconductivity can be induced
by the proximity to a metallic superconductor. Already, significant progress has been
made in fabricating highly transparent contacts between a graphene monolayer and a superconductor,
TiAl bilayer is proved to be a suitable material because of good electrical contact of Ti to graphene  (see
\textit{e. g.} \cite{grapheneSuperExper}). For the proximity induced superconductivity
$T_c$ is determined by that of the contacted superconductor
(for TiAl bilayer it is about 1.3 K). A superconductor-bilayer-graphene-superconductor setup
\cite{BilayerSuperConds}, with several S electrodes contacting
independently to the top and bottom layers can provide the
possibility for an induction of controllable interlayer
superconducting correlations. We anticipate that a hybrid system
(like a SNS Josephson junction) made of chiral superconductors to exhibit peculiar
properties due to the exotic order parameter and thermodynamic characteristics
of these type of superconductors. This can be used
for the experimental observation of chiral superconductivity, which may also provide potential
for new applications. It is worth to note that in both of the above proposed methods,
measurements sensitive to interlayer pairing, similar to those of the intrinsic Josephson effect
in layered high-$T_c$  superconductors(see \textit{e. g.} \cite{htctheorexp}), are required.
\par
In conclusion, we have predicted a new type of gapless
superconductivity which can be realized by an interlayer pairing of
relativistic-like chiral electrons in graphene materials. This chiral
superconductivity posses an exotic s-wave spin-triplet order parameter with anomalous
thermodynamic properties. We have found that the pairing gap can increase with
increasing of temperature, with the possibility of vanishing at a lower temperature
range including $T=0$, while having a finite value at higher temperatures. The entropy
of these anomalous stable superconducting states is found to be higher than that
of the normal state.

We acknowledge fruitful discussions with R. Asgari and M. R. Kolahchi.
We also gratefully acknowledge support by the Institute for Advanced Studies
in Basic Sciences (IASBS) Research Council under grant No. G2010IASBS110.
M. Z. thanks MPIPKS in Dresden and ICTP in Trieste for
the hospitality and support during his visit to these Institutes.

\end{document}